\documentclass[%
reprint,
 amsmath,amssymb,
 aps,
]{revtex4-2}

\usepackage{graphicx}
\usepackage{subfigure}
\usepackage{dcolumn}
\usepackage{bm}
\usepackage[dvipsnames]{xcolor}
\usepackage{multirow}
\usepackage{times}

\usepackage{hyperref}

\begin{document}
\hbadness=99999
\preprint{APS/123-QED}

\title{
Velocity dispersion of dark matter deficit ultra-diffuse galaxies: A case for modified gravity}

\author{Esha Bhatia}
\email{b.esha@iitg.ac.in}
\author{Sayan Chakrabarti}
\email{sayan.chakrabarti@iitg.ac.in}
\author{Sovan Chakraborty}
\email{sovan@iitg.ac.in}
\affiliation{Department of Physics, Indian Institute of Technology, Guwahati 781039, India}


\date{\today}

\begin{abstract}
 
The line of sight velocity dispersion of the ultra-diffuse galaxies (UDGs) NGC1052-DF2 and NGC1052-DF4 have been reasonably explained only with the baryonic matter, without requiring any dark matter contribution.
The comparable ratio between the baryonic and halo mass also ascertain the above claim for the two dark matter deficit galaxies. This paves the way for analyzing alternative gravity theories such as the $f(R)$ gravity and the Renormalization Group correction to General Relativity (RGGR). The analysis of the line of sight velocity dispersion shows that the choice of $f(R)$ gravity models such as Taylor expanded $f(R)$ about $R=0$ or a simple power law model of choice $R^n$ is consistent with the observational data. Similar statistical analysis is done for the RGGR and is also found to be a viable explanation for the observed velocity dispersion. We perform a global fit of the model parameters together with both the UDGs. The coupling parameters of the theories are considered as the global ones, and local variables such as the scale parameters are considered to be dependent on the individual galaxy.
\end{abstract}

\maketitle

\section{\label{intro}Introduction}
\setlength{\parskip}{6pt}
The affinity of the $\Lambda CDM$ \cite{Carroll:2000fy, 1984ApJ...284..439P} with the observations makes it a successful model to understand the dynamics of the Universe. Together with the introduction of Dark Matter (DM) and Dark Energy (DE) to the total energy budget of the Universe, $\Lambda CDM$ can explain the observations on both the cosmological and astrophysical scales, like accelerated expansion of the Universe \cite{SupernovaSearchTeam:1998fmf,weinberg2000cosmological}, missing mass problem in galaxies and clusters \cite{Weinberg:2013aya,Young:2016ala} etc. However, $\Lambda CDM$ also has certain shortcomings. On the astrophysical scales there are issues such as core-cusp problem, missing satellite problem \cite{Perivolaropoulos:2021jda,DelPopolo:2016emo}, satellite plane problem \cite{Pawlowski:2021ipt} and problems on large scales include fine tuning problem, horizon problem, coincidence problem and tensions between early and late time observations \cite{Bull:2015stt,Shankaranarayanan:2022wbx}. This motivates the study of  alternative proposals which introduces additional terms in the gravity action, for example adding higher order Ricci scalar, dynamical scalars, vectors or tensor fields, etc. \cite{Clifton:2011jh}. This in turn introduces extra degrees of freedom. In context of such alternative theories, the dynamical nature of Universe is explained by the modified action of gravity whereas the structure of energy momentum tensor constitutes only of baryonic matter. Such an alternative theory of gravity must be able to explain the gravitational dynamics both in the strong and weak-field regimes. In particular, it should regain a valid Newtonian limit on the Solar System scales. Additionally, it must be relativistic, self-consistent and a complete theory free of ghosts and instabilities. These above mentioned criteria are crucial to construct any consistent theory of gravity \cite{will_1993}. In this regard, we look into the kinematics of two such galaxies in different alternative gravity scenarios. The kinematics of the galaxies NGC1052-DF2 and NGC1052-DF4 are reasonably explained with the baryonic matter with almost no requirement for DM. The DM deficit nature of ultra-diffuse galaxy NGC1052-DF2 (hitherto mentioned as DF2) \cite{vanDokkum:2018vup} became a key as it was the first galaxy observed to have DM decoupled with baryonic mass. This contradicted the common structure that galaxies are enclosed within a spheroidal DM halo. Similar was the case with NGC1052-DF4 (hitherto mentioned as DF4) \cite{van_Dokkum_2019} which was also found to be DM deficit. Both the ultra diffuse galaxies (UDGs) are near the same host galaxy NGC1052 \cite{1978MNRAS.183..549F} and the DM deficit nature of galaxies came into light recently using Dragonfly Telescope Array (DTA) \cite{Abraham:2014lfa}. Thus, the kinematics of these dark matter poor UDGs become a promising tool to study the alternative gravity frameworks as the DM model details are of no consequence here. Indeed, the Newtonian dynamics alone without DM can explain the kinematics successfully. Thus, these systems can be considered to give the most conservative estimate on the alternative theory of gravity. Among the many gravity models, Milgromian Dynamics (MOND) \cite{Banik:2021woo} with external field effects coming from the host galaxy NGC1052 of DF2 and DF4 could accommodate the observed velocity dispersion (VD) \cite{Haghi_2019,Famaey:2018yif,Islam:2019irh,Islam:2019szu}. Similarly, VD evaluated for Scalar-Tensor-Vector Gravity (STVG), Emergent and Weyl Conformal gravity model was found to lie within the range of observations for UDGs \cite{Moffat:2018pab,Mannheim:2005bfa,Mannheim:1988dj,Dutta:2018oaj}. In the following, we study the kinematics of these UDGs in the light of the modified gravity models and find the global best fit parameters treating both the UDGs together. In particular the radial variation of the VD for these UDGs are analyzed for two different kinds of gravity models viz. $f(R)$ and Renormalization Group correction to General Relativity (RGGR) gravity.\\
\newline
The $f(R)$ gravity model  studied in this work assumes an arbitrary functional form of Ricci scalar ($R$) as compared to a linear $R$ dependence as is done in General Relativity (GR). The $f(R)$ model has a potential to give an alternative description to dark energy \cite{Nojiri:2006ri,Nojiri:2017ncd}. The model is also successful to explain inflationary era and bouncing cosmology \cite{Nojiri:2017ncd}. Many different functional forms for $f(R)$ have been proposed in the literature \cite{Clifton:2011jh,Nojiri:2006ri,Nojiri:2017ncd,Sotiriou:2008ve}. We study two such functional forms of $f(R)$ as motivated from \cite{Capozziello:2009vr,Capozziello:2006ph}. In general, the $f(R)$ theory is free of Ostrogradsky ghosts \cite{Woodard:2006nt, Motohashi:2014opa} which haunts the stability of many alternative gravity models. For $f(R)$ theory to be called successful it needs to be satisfied on all scales i.e, near and far IR regions. Our aim in this work is to test both the UDGs in context of two different $f(R)$ gravity models.\\
\newline
The first model assumes a more general functional form for $f(R)$ \cite{Capozziello:2009vr}. The solution for the weak-field potential for this model adds a Yukawa like term (exponential form) to the Newtonian potential. The exponential nature found here is not unique and is also observed and studied for other gravity models such as Scalar-Tensor Vector gravity \cite{Moffat:2014pia,Brownstein:2005zz}, Horndeski gravity \cite{Amendola:2019laa} etc. The earlier works on this model check the validity of the theory in regions of spiral and elliptical galaxies \cite{Capozziello:2009vr,Napolitano:2012fp}. The second functional form we analyze is motivated from  \cite{Capozziello:2006ph}. This model assumes a power law form for $f(R)$ i.e $f(R) \propto R^n$ with $n>1$ as a lower limit to the slope of the functional form. The Rotation Curve (RC) for a selection of Low Surface Brightness (LSB) galaxies (expected to be DM dominant) showed a promising fit with the observations for this power law model \cite{Capozziello:2006ph,Capozziello:2006uv,2014IJMPD..2342005S}. \\
\newline
We also look into another alternative gravity model which involves the quantum group correction to GR namely, Renormalization Group corrected General Relativity (RGGR) \cite{Rodrigues:2009vf}. The dependence of gravitational coupling constant $G$ on the energy scale can cause significant influence on the dynamics of the Universe \cite{Reuter:2004nx}. The RGGR model with the running of the gravitational coupling is parameterised in terms of a phenomenological parameter $\bar{\nu}$, whose variation can be as small as 10$^{-7}$. However, this small change can have impact on the kinematics of the galaxy. Previous works done on the RGGR model \cite{Rodrigues:2009vf,Reuter:2004nx,Farina:2011me,Rodrigues:2012wk,Rodrigues:2012wk,Rodrigues:2014xka} studied on astrophysical scales show that the kinematics can be satisfactorily explained. Interestingly, it is to be noted that the model parameter is found to have a linear dependence on the baryonic mass of the system \cite{Rodrigues:2012qm}\\
\newline
In this work, we focus on statistical analysis of the velocity dispersion data of these UDGs with and without (Newtonian only) the alternative gravity models. Our work statistically analyse the two $f(R)$ proposals  with the observed velocity dispersion data. Similar analysis is also done for RGGR gravity. We in this work initiate the probe of the said alternative gravity models with the velocity dispersion analysis. Additionally, we distinguish the local and global parameters of the alternative gravity models. For example, both the $f(R)$ proposals require scale parameters constrained by the size of the galaxy, such model parameters are treated as local ones. However, the coupling parameters in these models can be considered to be global and not dependent on the characteristics of the particular galactic system. Therefore, we initiate the global analysis of the alternative gravity models and fit the global parameters  together with the DF2 and DF4 VD observations data. We compare the results between the individual and global analysis.\newline 
\\
It is to be noted that our choices of the gravity models in this analysis are consistent \cite{will_1993, Shankaranarayanan:2022wbx}. However, these models involve free parameters which are varied in the consistency limits and are constrained from the observational data. The phenomenological models we probe are in contrast with the MOND which has a single fundamental constant in the theory. Contrary to MOND, 
 these gravity models contain parameters which are not fundamental but the theories are consistent according to the criterion mentioned above. 
\\
\newline
The paper is organized as follows. Section (\ref{form}) talks about the formalism where we discuss the analytical measurements to evaluate VD. Section (\ref{obs}) discusses the observations and density profile of UDGs DF2 and DF4. The next Section (\ref{model}) explains the different gravity models chosen to be looked into for such DM deficit galaxies. In section (\ref{res}), we describe the statistical analysis and compare the results. This section also mentions the fitting for individual and global scenarios. Finally, we conclude in Section (\ref{conc}).

\begin{figure*}
    \begin{subfigure}
    \centering
        \includegraphics[width=0.49\textwidth]{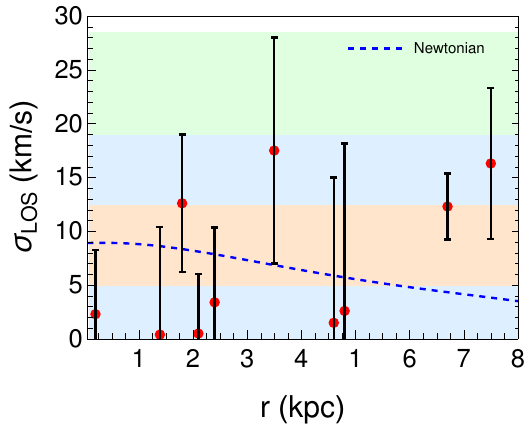}
    \end{subfigure}
    \begin{subfigure}
    \centering
        \includegraphics[width=0.49\textwidth]{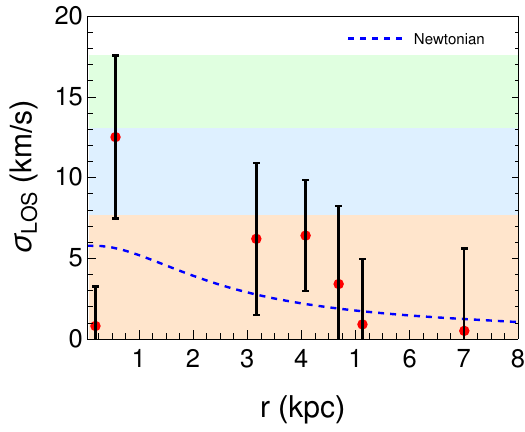}
    \end{subfigure}
\caption{\label{fgnw_udg} Radial variation of LOS VD for DF2 (left) and DF4 (right) shown using blue dashed line, assuming the underlying gravity to be Newtonian with no DM component. The red dots with the error bars are the observational VD measurements for individual globular cluster present within both the galaxies \cite{vanDokkum:2018vup,van_Dokkum_2019}. Additionally, we show the $1 \sigma$ (orange), $2\sigma$ (blue), $3\sigma$ (green) deviations of VD \cite{Haghi_2019} from observations.}
\end{figure*}

\section{\label{form}Formalism}
 Velocity Dispersion (VD) is an important tool to understand the kinematics of galaxies which are not rotationally supported  \cite{Wake_2012}. The VD ($\sigma$) of a star under the influence of gravitational potential of the surrounding mass of a galaxy is governed by the radial Jeans equation \cite{2008gady.book.....B};
\begin{equation}\label{sgma}
   \frac{ \partial}{\partial r}(\rho(r) \sigma^2(r))+\frac{2\rho(r)\beta\sigma^2(r)}{r}=\rho(r) a(r).
\end{equation}
Here we assume a spherically symmetric mass distribution and $r$ is the radial distance from the center of the galaxy, $\beta$ is the anisotropy parameter which is the measure of deviation of VD from radial isotropy, $a(r)$ is the acceleration due to gravity and $\rho(r)$ is the density profile for the galaxy. For an isotropic system the parameter $\beta = 0$, implying equal radial and the tangential components of VD. In this study we treat our systems to be isotropic and under this assumption the expression for VD reduces to,
\begin{equation}
    \sigma^2(r)=\frac{1}{\rho(r)}\int_r^{\infty}\rho(r')a(r')dr'.
\end{equation}
However, the physically relevant quantity measured in astrophysical observation is the projection of the above mentioned radial VD on the line joining the observer and the center of the system, termed as the the line of sight (LOS) velocity dispersion. Thus, $\sigma_{LOS}$ expressed in terms of the the radial VD and the density profile of the system is written as:
\begin{equation}\label{sgmalos}
    \sigma^2(R)_{LOS}=\frac{\int_R^{\infty} \frac{\sigma^2(r) \rho(r)}{\sqrt{r^2-R^2}}dr}{\int_R^{\infty} \frac{r\rho(r)}{\sqrt{r^2-R^2}}dr}.
\end{equation}
The LOS VD is a function of R, which is the projected distance from the center of the galaxy. Therefore, given the weak-field limit potential for a gravity model, one can compute the radial acceleration. This evaluates the radial VD for a given mass density ($\rho (r)$) for the galaxy. Now, the variation of the LOS VD with $R$ can be estimated and matched with the observational data which puts constraints on the parameters of different gravity models in question.
\section{\label{obs}Observations} 
Observed baryonic matter with GR as the underlying theory of gravity fails to explain the flattening of the galactic rotation curves at large radius for majority of the galaxies. One way to resolve the issue is the addition of DM together with the baryonic matter \cite{1980ApJ...238..471R}. This problem has also been studied by modifying the underlying theory of gravity with or without DM. The explanation with the existence of DM is supported from the perspective of cosmology and astro-particle physics. Thus, for such galactic systems, the phenomenology of modified gravity gets complicated due to the presence of additional DM model parameters. In this sense, any DM deficit system is unique and would provide a test bed for different modified gravity models. \\
The first galaxy discovered \cite{vanDokkum:2018vup} with negligible presence of DM was NGC1052-DF2  (here on mentioned as DF2). This ultra diffuse spheroidal galaxy DF2 is located near the host galaxy NGC1052 which was originally discovered in \cite{1978MNRAS.183..549F}. The total mass enclosed within the radius of $7.6$ kpc for DF2 is $3.4\times10^8M_{\odot}$ \cite{1981ApJ...244..805B}. The observed ratio $M_{DM}/M_{stellar}$ is of $\mathcal{O}(1)$ which is almost $400$ times lower than typical galaxies enclosed by DM halos \cite{vanDokkum:2018vup}. The radial velocity of the $10$ globular clusters within DF2 were also studied from the Doppler shift of Calcium triplet line \cite{vanDokkum:2018vup}. These observations suggest an average velocity of $1803$ km/s for these globular clusters. The observed VD at $90\%$ CL turns out to be less than $10.5$ km/s. The mass density of DF2 is  estimated from the luminosity distribution using the stellar mass to light ratio. The luminosity distribution is parameterized using the Sersic profile   \cite{Graham:2005fy, 2008gady.book.....B} 
with Sersic index $n=0.6$, axis ratio $b/a=0.85$ and half-light radius as $2.2$ kpc. 
The stellar mass for the galaxy calculated from this density profile is $2\times10^8$ M$_{\odot}$
and is consistent with the low DM interpretation of DF2 \cite{vanDokkum:2018vup}. Another UDG observed \cite{van_Dokkum_2019} within the same group NGC1052 i.e, NGC1052-DF4 (here on mentioned as DF4) shows similar nature. The spectroscopy and imaging confirmed the presence of 7 globular cluster like objects within. Similarly for DF4 the luminosity is estimated by the Sersic profile with Sersic index $n=0.79$, axis ratio $b/a=0.89$ and major-axis half light radius  $1.6$ kpc. Similar analysis for DF4 showed intrinsic VD measurement of about $4.2^{+4.4}_{-2.2}$ km/s. 
 The ratio of M$_{DM}$/M$_{stellar}$ computed from the VD observation is of the $\mathcal{O}(1)$, suggesting the existence of a second galaxy lacking DM \cite{van_Dokkum_2019}.\\
\newline
The density profiles for both the UDGs, DF2 and DF4 are found to be similar and can be approximated as \cite{Moffat:2018pab};
\begin{equation}\label{den}
    \rho(r)\sim \frac{40 \rho_0}{63 r_s}exp\left( -\left[\frac{11r}{10r_s}\right]^{4/3}\right),
\end{equation}
where $\rho_0$ is the characteristic surface mass density and $r_s$ is the effective radius. 
Both the ultra-diffuse galaxies are at a distance of about $20$ Mpc \cite{vanDokkum:2018vup,van_Dokkum_2019}, resulting in the density parameters ($\rho_0$, $r_s$) for DF2 to be ($1.25 \times 10^7$ M$_{\odot}$/kpc$^2$, $2$ kpc) and (1.15 $\times$10$^7$ M$_{\odot}$/kpc$^2$, 1kpc) for DF4.
Given these density profiles one may estimate the VD for the system.

To begin with, we probe the standard gravity paradigm described by Newtonian gravity from the observational data for the two UDGs.  The acceleration for a spherically symmetric system in Newtonian gravity is written as;
\begin{equation}\label{nw_ac}
    a(r)=-\frac{G}{r^2}\int_0^r 4\pi\rho(r')r'^2dr' .
\end{equation}
Here, $G$ is the gravitational constant and $\rho(r)$ is the approximate density distribution of UDG as given in eq.(\ref{den}). The LOS VDs evaluated for both the galaxies are shown in Fig.\ref{fgnw_udg}. The left panel is for DF2 and right one shows the case for DF4. The blue dashed lines in both panels show the estimated radial variation of LOS VD. Additionally, the orange, blue and green shaded regions are the $1\sigma$ and $2\sigma$ and $3\sigma$ regions on the predicted VD measurements \cite{Haghi_2019}. The red dots with error bars are the observations of the individual globular objects present within the galaxy i.e, 10 such objects for DF2 \cite{vanDokkum:2018vup} and 7 objects for DF4 \cite{van_Dokkum_2019}.

The above mentioned UDGs due to the lack of DM are best candidates to probe the parameter space for different alternative gravity models, resulting in LOS VDs different than the Newtonian curves shown in Fig.\ref{fgnw_udg}. In particular, for the modified gravity in the weak-field limit, $a(r)$ will have an additional term compared to the Newtonian scenario. We use this modified expression for acceleration to estimate the velocity dispersion and constrain the gravity model parameters from the UDG observations.

\section{\label{model}Models} 

 The breakdown of GR at different length scales has led to the proposal of many alternative models of gravity. One set of models where the Ricci scalar $R$ in the Einstein-Hilbert (E-H) action is replaced by an analytic function of the $R$, known as the $f(R)$ gravity models, has
 been extensively studied in the context of astrophysics  and cosmology. As already mentioned earlier, one of the crucial successes of using such $f(R)$ models was to explain the cosmic acceleration. It may be noted at this point that any analytic form of $f(R)$ may give rise to a new model of gravity, but it is crucial to get the model tested against observations. Towards this direction, testing the $f(R)$ theories against observations at astrophysical as well as cosmological scales is extremely important. For example, on the astrophysical scales, the velocity dispersion observations for the UDGs can be used to constrain the modified gravity parameters. In our work, we plan to study the kinematics of the two UDGs by choosing two different types of gravity models viz., $f(R)$ gravity and Renormalization Group corrected General Relativity (RGGR). Validity of these gravity models are probed by scanning the free model parameter space against the UDG velocity dispersion observations. 

\subsection{\texorpdfstring{$f(R)$}{TEXT}  gravity}

In an alternative approach to $\Lambda$CDM, we can assume E-H action to have a general $f(R)$ form instead of Ricci scalar $R$ \cite{Nojiri:2006ri, Faraoni2008fRGS}. 
The literature suggests many different functional forms for $f(R)$ which satisfies different observational regimes of the Universe \cite{Clifton:2011jh}. These different choices for the functional form add to the complexity of the theory. Additionally, there can be models such as $f(R)\propto 1/R$ which consistently explains the acceleration of the Universe without the need for DE but suffers from an instability as discussed by Dolgov and Kawasaki \cite{Woodard:2006nt}.\\
\newline
Keeping the above mentioned factors in mind we look into two cases of $f(R)$. The first model assumes a generalized functional form which is Taylor expanded about a flat background. Similarly, for the second model a specific form of $f(R)$ that is $\propto R^n$ is chosen and analyzed.


\subsubsection{\texorpdfstring{$f(R)$}{TEXT} gravity (model A)}
This generic choice of $f(R)$ gravity model assumes a Taylor expansion of the functional form about the Minkowskian background  that is $R=0$ \cite{Capozziello:2009vr,Lubini:2011pc}:
\begin{equation}
    f(R)\simeq \sum_{i=0}^{\infty} \frac{f_i(0) R^i}{i!},
\end{equation}
where $f_i$ are the coefficients associated with the expansion. The first term of the series is a constant and can be set to zero. Thus the further solution has no dependence on the parameter $f_0$. The resultant solution in the weak-field limit can yield a modified potential given as \cite{Capozziello:2009vr}:
\begin{equation}\label{yuk_fr}
\phi(r)=-\left(\frac{GM}{1+\delta}\right)\frac{1+\delta e^{-r/\lambda}}{r} .
\end{equation}

In the above eq.(\ref{yuk_fr}), the additional Yukawa like term is characterized by $\delta$ and $\lambda$. Also, as can be seen from the equation above, for the point like baryonic mass $M$ the Newtonian part is scaled by a factor $\frac{1}{1+\delta}$. 
Here, $\delta$ is the coupling which determines the nature of the additional force and can be attractive or repulsive depending on the sign of the parameter. Following \cite{Stabile:2013jon}, we consider $\delta$  to be negative and within the range $-1<\delta<0$, implying repulsive nature of Yukawa force. Substituting $\delta=0$ in eq.(\ref{yuk_fr}), gives back the GR case where potential varies as $1/r$. The parameter $\lambda$ is the scale length and corresponds to the size of the system under analysis. Thus, $\lambda$ is not an universal parameter. However, this $f(R)$ theory still satisfies the consistency conditions to be a valid theory of gravity.
This gravity model has been looked into different regions ranging from galactic clusters dynamics \cite{Capozziello:2008ny} to studying the VD of elliptical galaxies \cite{Napolitano:2012fp}. The analysis shows that the Yukawa model has the potential to explain the kinematics on astrophysical scales without any need for DM. The phenomenological parameter $\delta$ obtained lies roughly within the range ($-0.7,-0.9$). The scale parameter $\lambda$  is also found to be dependent on the size of the systems.
\newline

\subsubsection{\texorpdfstring{$f(R)$}{TEXT} gravity (model B)}
This $f(R)$ model replaces the Ricci scalar ($R$) in the action by a power-law form given as $f_0 R^n$ \cite{Capozziello:2006ph}. Solving the equation of motion in the weak-field limit for the spherically symmetric system yields the potential,
\begin{equation}\label{yukp}
    \phi(r)=-\frac{Gm}{2r}\left[1+\left(\frac{r}{r_c}\right)^{\beta}\right],
\end{equation}
where $G$ is the Newtonian gravitational constant, $r_c$ is the scale radius and $\beta$ is related to the power ($n$) of the model as
\begin{equation}\label{btrc}
    \beta=\frac{12n^2-7n-1-\sqrt{36n^4+12n^3-83n^2+50n+1}}{6n^2-4n+2}.
\end{equation}
Under the conditions that the gravity is Newtonian on the Solar system scales and the potential converges even at large distances 
$\beta$ is constrained in the range $0 <\beta< 1$ \cite{Capozziello:2006ph}. Additionally, $r_c$ is the scale radius at which the modified gravity takes into effect. Thus, the parameter $r_c$ is not universal in nature as it can change with the scale of the problem. However, this alternative gravity model still fulfils all the criteria to be a consistent theory.
For example, on the astrophysical scales the rotation curves of Spiral galaxies are reasonably explained with the above gravity model, competitive to the $\Lambda CDM$ explanation \cite{2014IJMPD..2342005S}. Similarly, study  of rotation curve of LSB galaxies showed that a global value $n=3.5$ or $\beta =0.85$ can sufficiently explain the circular velocities with $r_c$ being a local variable depending on scale of the system \cite{Capozziello:2006ph,Capozziello:2006uv}.
\begin{figure*}
    \begin{subfigure}
        \centering
         \includegraphics[width=0.49\linewidth]{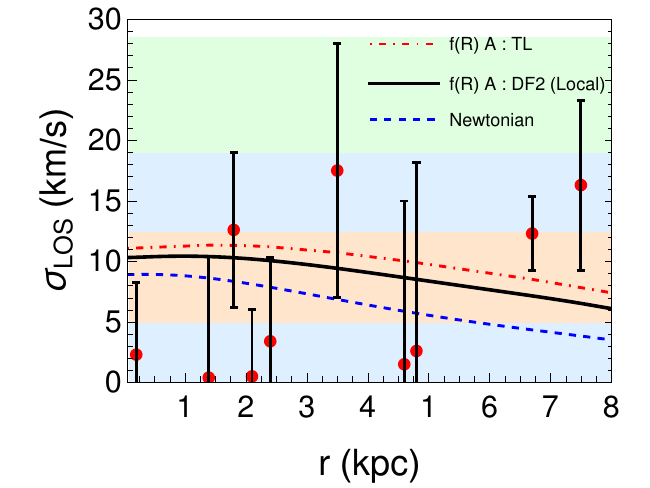}
    \end{subfigure}
    \hfill
    \begin{subfigure}
        \centering
         \includegraphics[width=0.49\linewidth]{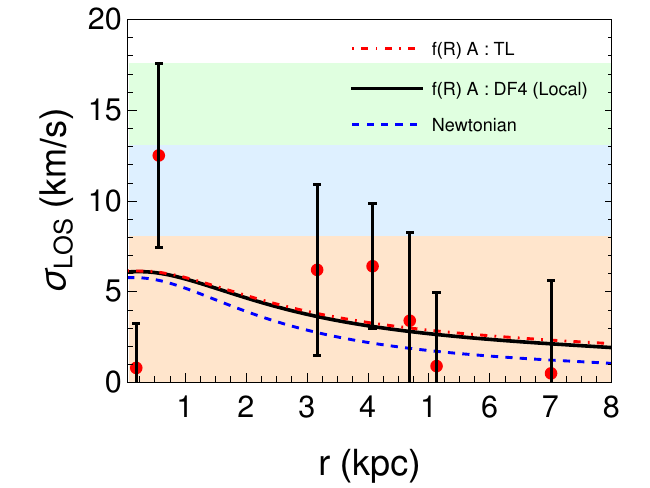}
    \end{subfigure}
    \caption{\label{yuk_f24}Comparison of the Newtonian radial variation of VD (blue dashed curve) for both the UDGs with the $f(R)$ gravity model A. The black solid line is the velocity dispersion for $f(R)$ gravity model A from individual likelihood analysis of the DF2 (left) and DF4 (right). The red dot-dashed lines depict the best fit LOS VD from the total likelihood (TL) analysis. Other details of the plots remains same as given in Fig.\ref{fgnw_udg}}.
\end{figure*}
\subsection{RGGR}
In this model we incorporate the Renormalization Group corrections to General Relativity (RGGR) following \cite{Rodrigues:2009vf}. 
It is well known in quantum field theory that the relation between the coupling constant and
the energy scale is given by a beta function,
$ \beta_i=\mu \frac{dg_i}{d\mu}$, where $g_i$ are the effective coupling constants and $\mu$ is the RG scale. Therefore, by knowing the beta function, the coupling functions $g_i(\mu)$ can be easily determined. Although, the above approach works well in case of field theories in flat space-time, the case of gravity is different. The theory arising out of the Einstein-Hilbert action is a non-renormalizable one and at present there is no complete theory of quantum gravity. Nonetheless, some steps can be taken towards renormalizing GR by treating it as a field theory in curved space-time (see \cite{Shapiro:2008sf} for a review and related references). In the renormalization approach, the gravitational coupling $G$, which is constant in the far infrared regime (IR) is assumed to vary with the energy scale ($\mu$). Note that, there is no proof that the RG induced running of the gravitational constant $G$ becomes zero at very low energies/large scales, however, if the gravitational coupling  runs, a natural beta function as predicted in the literature \cite{Reuter:2004nx, Farina:2011me} in various different contexts is estimated as $\beta=\mu\frac{dG^{-1}}{d\mu}$. In particular, the variation of gravitational coupling $G$ is taken to be \cite{Rodrigues:2014xka}
\begin{equation}
    \mu \frac{dG^{-1}}{d\mu}=2\nu\frac{M_{Planck}}{c^2\hbar}=2\nu G_0^{-1},
\end{equation}
Solving the above equation, the variation of $G$ with energy scale takes the form:
\begin{equation}\label{energy}
    G(\mu)=\frac{G_0}{1+\nu ln \frac{\mu^2}{\mu_0^2}},
\end{equation}
where $\mu_0$ is the reference energy scale such that $G(\mu_0)=G_0$ and $G_0$ is the measured value of gravitational constant in Solar System. Also, $\nu$ is a phenomenological parameter determining modification to the gravitational coupling due to RG effect. GR can be recovered by substituting $\nu$ = 0 in eq.(\ref{energy}) and gravitational coupling becomes a constant parameter. The coupling parameter is sensitive to $\nu$ hence, a small variation of the order of $10^{-7}$ can make a significant difference on the galactic scales.  Additionally, correlating the energy scale with some observable i.e, Newtonian gravitational potential ($\phi_N$) results in the relation \cite{Domazet:2010bk};
\begin{equation}
    \frac{\mu}{\mu_0}=\left(\frac{\phi_N}{\phi_0}\right)^{\alpha},
\end{equation}
where $\phi_0$ is the potential measured in the Solar system 
and $\alpha$ is the free parameter linearly dependant on the mass of the galaxy \cite{Rodrigues:2012qm}. Now, the E-H action of gravity contains, in addition to the Ricci scalar, an energy dependent coupling parameter $G$. Studying the equation of motion in the weak-field limit, the circular velocity takes the form:
\begin{equation}\label{rgr_v}
    V^2(r)=V_N^2(r)\left(1-\frac{\bar{\nu}c^2}{\phi_N(r)}\right),
\end{equation}
where $V_N$ and $\phi_N$ are the Newtonian circular velocity and potential respectively and $c$ is the speed of light. 
The two unknown phenomenological parameters ($\nu$, $\alpha)$ can be written as a single unit i.e, $\bar{\nu} \equiv \nu \alpha$. The connection of the parameter with the stellar mass can be understood from the example of Solar system having baryonic mass $10^{-10}$ times lesser than a typical galaxy. Thus, the phenomenological bound obtained (from solar system ) on $\bar{\nu}$ is $\leq 10^{-17}$ which is consistent with the mass dependence claim of the parameter \cite{Farina:2011me}. The dependence of the phenomenological model parameters on mass or running of gravitational coupling parameter with defined energy scales suggests that the 
free parameter changes with the scale of the problem and hence not universal in nature. However, the model still satisfies the  consistency conditions to remain a valid choice. In fact, phenomenological study for a selection of galaxies \cite{Rodrigues:2014xka} finds the $\bar{\nu}$ to have a nearly linear relation with the baryonic mass. In general, the acceleration on the galactic scales can be written as:
\begin{equation}\label{argr}
    \phi'_{RGGR}(r)\approx\phi_N'(r)\left( 1-\frac{c^2\bar{\nu}}{\phi_N(r)} \right),
\end{equation}
where $\phi'_{RGGR}(r)$ and $\phi_N'(r)$ is the RGGR and Newtonian acceleration due to gravity respectively.
For the spherically symmetric galaxy such as UDGs, $\phi_N$ takes the  form:
\begin{equation}
 \phi_N(r)=-\left(\frac{GM(r)}{r}+4\pi G \int_r^{\infty}\rho(R)R dR\right),
\end{equation}
In the above equation $M(r)$ is the stellar mass contained within the radius $r$ and $\rho(r)$ is the density profile of the galaxy given in eq.(\ref{den}). The Newtonian potential $\phi_N(r)$ also takes into account the effect of mass present external to the radius $r$. \\
\newline
The phenomenological study of the RGGR model on elliptical and disk galaxies shows best fit value of $\bar{\nu}$ lies in the range of $10^{-6}-10^{-8}$. The results obtained were in favour with the observed dynamics of the galaxies that were looked into \cite{Rodrigues:2012qm,Farina:2011me}.

\begin{table}[ht]
\begin{tabular}{|p{0.23\linewidth}|p{0.18\linewidth}|*{4}{c|}} 
\hline
\bf Gravity model & \bf Model parameters & \multicolumn{2}{c|}{\bf NGC1052-DF2}& \multicolumn{2}{c|}{\bf NGC1052-DF4} \\ 
\hline
 &&Local&Global&Local&Global\\
\hline
\multirow{2}{6.32em}{\bm{$f(R)$}\bf~model A} & $~~~~~~\delta$ & $-0.82$ & $-0.89$ & $-0.83$ & $-0.89$\\
 &~~~~ $\lambda$ (kpc) & $~~~6.16$ & $~~~6.80$ & $~~~5.39$ & $~~~6.50$\\
\hline
\multirow{2}{6.3em}{\bm{$f(R)$}\bf~model B} & $~~~~~~\beta$ & $~~~0.65$  & $~~~0.60$ & $~~~0.50$ & $~~~0.60$ \\
& $~~~~r_c$ (kpc) & $~~~0.30$ & $~~~0.40$ & $~~~0.25$ & $ ~~~0.35$\\
 \hline
\multirow{1}{5em}{\bf~~~~ RGGR} & $~~~~\bar{\nu}$ $\times$ $10^{-8}$ & $~~~0.30$ & $~~~0.16$ & $~~~0.10$ & $~~~0.16$ \\
\hline
\end{tabular}

\caption{The model parameters for different alternative gravity models fitted for DF2 and DF4 are shown above.   The local columns in the table contains best fit values for the analysis with the individual galaxy likelihood. The global column fit values are for the total likelihood. For the global analysis the parameters $\delta$, $\beta$ and $\bar{\nu}$ in their respective models are considered as global and the parameters $\lambda$ and $r_c$ remained local to individual galaxies.}
\label{Tab:mogpar}
\end{table}
\begin{figure*}
    \begin{subfigure}
        \centering
         \includegraphics[width=0.49\linewidth]{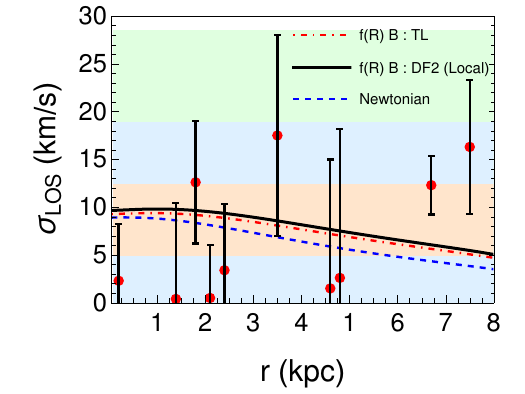}
    \end{subfigure}
    \begin{subfigure}
        \centering
         \includegraphics[width=0.49\linewidth]{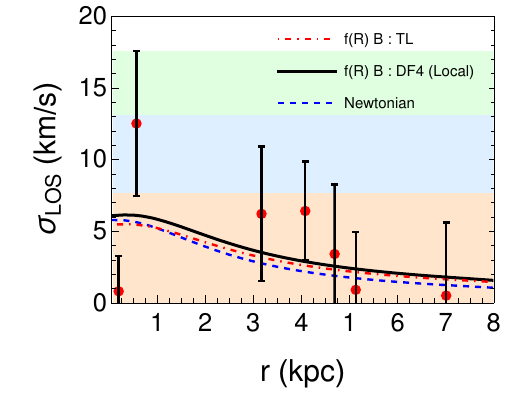}
    \end{subfigure}
    \caption{\label{cap_f24} Comparison of the Newtonian radial variation of VD (blue dashed curve) for both the UDGs with the $f(R)$ gravity model B. Radial VDs for the $f(R)$ model are shown by solid black lines when both galaxies i.e, DF2 (left) and DF4 (right) are treated independently. The VDs from the global analysis using total likelihood (TL) with both the UDGs are shown using the red dot-dashed lines. Other information about the Newtonian contribution and observations in the plots remain the same as in Fig.\ref{fgnw_udg}}.  
\end{figure*}
\section{\label{res}Results}
To study the dynamical behaviour of both the UDGs with respect to the alternative gravity models we evaluate the model parameters by comparing the estimated VD with the observational data. We perform the statistical analysis in two different ways. The first scenario constrains all the model parameters for both the galaxies independently, this we interpret as the local estimate. For the other scenario, the global parameters for a model are fitted from both the UDG analyzed together. While the local parameters, like the scale parameters, are fitted individually for each UDG.
To constrain the model parameters we define a likelihood function for each galaxy under the assumption that errors follow a Gaussian distribution \cite{deAlmeida:2018kwq}:
\begin{equation}\label{chi}
\resizebox{0.98\hsize}{!}{
    $\mathcal{L}_j(\bm{\theta},\bm p)= (2\pi)^{-N/2} \left\{\displaystyle \prod_{i=1}^N\sigma(r_i)^{-1}\right\} \exp\left\{-\frac{1}{2} \displaystyle\sum_{i=1}^{n}\left(\frac{\sigma_{obs}(r_i)-\sigma_{LOS}(r_i,\bm{\theta},\bm p)}{\sigma_{err}(r_i)}\right)^2\right\}$,}
\end{equation}
where $\sigma_{obs}$ are the observational data points for VD and $\sigma_{LOS} (r_i,\bm{\theta},\bm p)$ is estimated for the parameters  ($\bm{\theta}$, $\bm p$) pertaining to the alternative gravity models. In this definition, the parameters $\bm{\theta}$ represent the global ones, while  $\bm p$\ is solely dependent on the individual galaxy property and are the  local parameters of the gravity model. Here, $\sigma_{err}$ is the error from the observational data.  For the $f(R)$ gravity the global parameters are $\bm{\theta}$=\{$\delta$\} and $\bm{\theta}$=\{$\beta$\} for model A and B respectively. The local parameters for model A and B are \{$\bm p$\}=\{$\lambda$\} and \{$\bm p$\}=\{$r_s$\}, respectively. Both these parameters depend on the size of galaxy. Contrary to the $f(R)$ model, RGGR gravity has a single mass dependent free parameter i.e, $\bm{\theta}$=\{$\bar{\nu}$\}. However the mass being similar for both the UDGs one may treat  the $\bar{\nu}$ as a global parameter in this study. 
To compute the global parameters we now define the total likelihood function. The observations for both the galaxies being independent of each other allow the total likelihood function to be defined as the product of the likelihood function for each galaxy:
\begin{equation}
    \mathcal{L}(\bm{\theta} ,\bm p)=\displaystyle \prod_{j=1}^{2} \mathcal{L}_j(\bm \theta,\bm p),
\end{equation}
The best fit value for the parameter $\bm{\theta}$ and $\bm p$ are found by the minimization of the log-likelihood function  with respect to the model parameters. 
\subsection{\texorpdfstring{$f(R)$}{TEXT} gravity}
\subsubsection{\texorpdfstring{$f(R)$}{TEXT} (model A):}
The parameter $\delta$ for the $f(R)$ model A is interpreted as the coupling signifying the deviation from the Newtonian gravity and  assumed to be in the range ($-1,0$). The other parameter $\lambda$ is the scale radius and varies within the range of size of the UDGs. The optimized model parameters values are estimated using likelihood minimization. 
The best fit ($\delta$, $\lambda$) obtained for individual analysis of DF2 and DF4 are ($-0.82$, $6.16$ kpc) and ($-0.83$, $5.39$ kpc), respectively. These results are shown as the local fit parameters in Table \ref{Tab:mogpar}. 
As both the UDGs (DF2 and DF4) have almost similar mass and size, the coupling and scale parameter as seen from Table \ref{Tab:mogpar} are also found to be similar. For  the global analysis incorporating both UDGs, $\delta$ is treated as the global parameter while the scale dependence parameters ($\lambda_{DF2}, \lambda_{DF4}$) vary independently for each galaxy. The global fit value for coupling parameter $\delta$ turns out to be $-0.89$ and $\lambda_{DF2} = 6.8$ kpc,  and  $\lambda_{DF4} = 6.5$ kpc. These fit parameters are shown as global in the Table \ref{Tab:mogpar}. Comparing the global parameter from UDGs with the study of elliptical galaxies \cite{Napolitano:2012fp} it is observed that $\delta$ lies within the expected range i.e, ($-0.7, -0.9$) from the previous phenomenological study. Also as different galaxies are of varying sizes, $\lambda$ is different but the order of the parameter does not changes drastically. This signifies that the study of DF2 and DF4 is consistent with the previous analysis done on the gravity model. \\
\newline
For these best fit parameters we plot the line of sight VDs in Fig. \ref{yuk_f24}. The DF2 results are shown in the left panel and DF4 are in the right one. The local LOS VD results are shown by the solid black curves in Fig.\ref{yuk_f24}. The red dashed lines in the plots represent the global scenario with the fit parameters 
optimized from the total likelihood. The Newtonian VD contribution is shown by the solid blue line. The $1\sigma$ and $2\sigma$ and $3\sigma$ regions on the predicted VD measurements \cite{Haghi_2019} are also shown by the orange, blue and green shaded regions. The red dots with error bars are the observations of the individual globular objects present within the galaxy.  Our estimated VDs for all the different fit scenarios come within the $1\sigma$ region. Comparing the LOS VDs in both galaxies we observe that the difference of the alternative model with Newtonian gravity is smaller in case of DF4 galaxy. In general, the VD plot shows that $f(R)$ gravity to be a feasible choice and can explain the kinematics for both the DM deficit UDGs. Results from both the individual likelihood analysis and total analysis are also in agreement.
\begin{figure*}[ht]
    \begin{subfigure}
        \centering
         \includegraphics[width=0.49\linewidth]{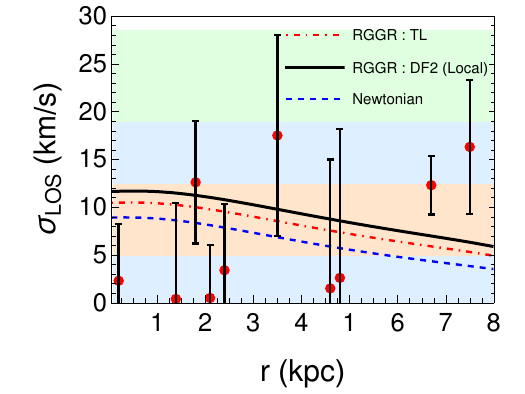}
    \end{subfigure}
    \begin{subfigure}
        \centering
         \includegraphics[width=0.49\linewidth]{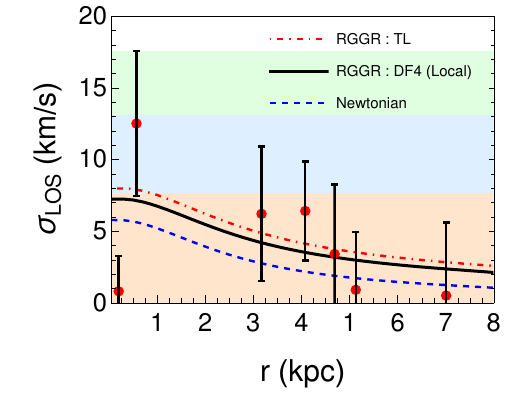}
    \end{subfigure}
        \caption{\label{rgr_f24} Comparison of the Newtonian radial variation of VD (blue dashed curve) for both the UDGs with the RGGR gravity. Radial VDs for the RGGR model are shown by solid black lines when both galaxies i.e, DF2 (left) and DF4 (right) are treated independently. The VDs for the global analysis from total likelihood (TL) for both the UDGs are shown using the red dot-dashed lines. Other information about the Newtonian contribution and observations in the plots remain the same as in Fig.\ref{fgnw_udg}}
\end{figure*}
\subsubsection{\texorpdfstring{$f(R)$}{TEXT} (model B):}
This $f(R)$ model is also described by two free parameters ($\beta$, $r_c$). The dimensionless parameter $\beta$ varied in the range ($0, 1$), has a quadratic relation with the power $n$ as given in Eq. (\ref{btrc}). The parameter $r_c$ can be interpreted as the local scale parameter varying within the size of the galaxy. The fitting parameters ($\beta$, $r_c$) obtained from the likelihood analysis are summarized in Table \ref{Tab:mogpar}.

For DF2 and DF4, the fit value ($\beta$, $r_c$) for the individual likelihood analysis obtained are ($0.65, 0.30$ kpc) and ($0.50, 0.25$ kpc) respectively. Correspondingly, the value of $n$ evaluated for DF2 and DF4 from eq.(\ref{btrc}) is $1.92$ and $1.46$, respectively. For the global analysis of the combined UDGs the $\beta$ obtained is $0.60$ ( $n = 1.72$) and the scale parameters obtained for DF2 and DF4 are $0.40$ kpc and $0.35$ kpc, respectively. The comparison of the global parameter $\beta$ from our study and the analysis of LSBs ($\beta=0.8$) \cite{Capozziello:2006ph} shows that the difference between the two is not large. Also, the scale parameter $r_c$ dependent on the mass of the galaxy is within the same range as obtained from the study of other LSBs. Thus, the kinematics of UDGs is consistent with the earlier work done on the $f(R)\propto R^n$ gravity model choice. \\
\newline
From the obtained best fit parameters, the radial variation of LOS VD is shown in Fig.\ref{cap_f24}. The black solid and red dashed line in Fig.\ref{cap_f24} are the local and global VD contribution for the $R^n$ gravity model for DF2 (left) and DF4 (right). The blue dashed lines in both the plots are the VD contribution when underlying gravity is taken to be Newtonian. For both the UDGs the black and red dashed line lies within $1\sigma$ region of the LOS VD observation. Thus, signifying that a model having a specific $R^n$ functional form for $f(R)$ can also be accommodated to explain the dynamics of our chosen UDGs.
Both the global and the local fits for this $f(R)$ model also are in agreement and signifies the similarity in the properties of the two UDGs.

\subsection{RGGR}
In the case of RGGR gravity, potential in the weak-field limit is a function of the parameter $\bar{\nu}$. The $\bar{\nu}$ parameter is found to have nearly linear dependence on the mass of galaxy. DF2 and DF4 having similar baryonic mass are expected to have similar values of $\bar{\nu}$. Therefore, for the global analysis we treat $\bar{\nu}$ as a single parameter and to be optimized for both the galaxies.
The individual likelihood analysis for both DF2 and DF4 evaluates $\bar{\nu}$ to be $0.30\times10^{-8}$ and $0.10\times10^{-8}$ as seen from Table \ref{Tab:mogpar}. Additionally, $\bar{\nu}$ is also constrained from the total likelihood for both UDGs and is found to be $0.16\times10^{-8}$. This small value of $\bar{\nu}$ parameter is correlated with the UDG mass. As both the UDGs have low mass in comparison to typical galaxies, the estimated $\bar{\nu}$ is atleast one order smaller to the previous analysis of spiral galaxies where $\bar{\nu}$ lies in the range $10^{-6}-10^{-8}$ \cite{Rodrigues:2014xka}.

The line of sight VDs for these likelihood fitting are shown in the Fig.\ref{rgr_f24}. Similar to the previous figures  the DF2 case is shown in the left panel and the DF4 results in the right panel. In both panels, the VDs for the local individual likelihood results are shown by solid black lines. Similarly, the LOS velocity dispersion curves for the global scenario are shown by the red dot dashed line. The LOS VD contribution from the Newtonian gravity is represented by the solid blue lines.\\
As seen from Fig.\ref{rgr_f24}, the RGGR gravity contributions from both the analysis lie within the $1\sigma$ region of the observations (orange region). The fact that the underling alternative gravity model of RGGR can explain the dynamical behaviour of both the UDGs remains true for both the global and local analysis.

\section{\label{conc}Conclusion}
Several proposals are made for the Einstein's action of gravity by introducing terms such as vector, scalar, tensor field in addition to Ricci scalar. Astrophysical observations are often used as a testing probe for these models. The free parameters of the alternative gravity models are constrained from observations on different length scales. In this regard, the discovery of UDGs such as DF2 and DF4 lacking DM paves the path to study such modified gravity models. The kinematics of these UDGs can be explained with normal stellar matter without invoking the need for DM. Hence, phenomenological study of the alternative gravity models are free from the DM properties. In this study, we have analyzed the line of sight VD for such galaxies for two alternative gravity models i.e, $f(R)$ gravity (a chosen $f(R)$ form and a general case) and RGGR gravity.\\
\newline
The free parameters of the gravity models have been estimated by the likelihood analysis. While, constraining the model parameters from both UDGs we treat certain parameters as a global one and find the fit from  both the galaxies. The case of $f(R)$ gravity is studied from two perspective. The first model chooses a functional form $f(R)$ and adds a Yukawa like term to the Newtonian potential. The second choice takes an effective power law form for the $f(R)$ resulting in a radius dependent power term ($\propto r^{\beta}$) to the Newtonian potential. Studying both the DF2 and DF4 galaxies for the two $f(R)$ scenarios shows that the best fit VD curve lies within the $1\sigma$ region of the observation. Similarly, for the case of RGGR model a potential-dependent term is added to the Newtonian gravity. The free parameter of this model increases in a nearly linear manner with mass of the galaxy. Hence, the free parameter ($\bar{\nu}$) for the two UDGs having almost similar mass turns out to be similar. The optimized VD fitted with the free parameters also lies within the 1$\sigma$ range of observations. Thus, all the alternative  gravity models in this discussion are found to be reasonable while explaining the VD data.\\'
\newline
In summary, the observed luminous content within both the UDGs sufficient to understand the kinematics. Thus, GR without DM is found to be  consistent with the observed VD for the galaxies. However, when the UDGs are looked into the light of our chosen alternative gravity models, the observed VD are found to be consistent with our choices of the alternative gravity. The analysis suggests that these gravity models with the present observational data precision cannot be ruled out.\\
\newline
However, the presence of large errors on the present observations for both the UDGs puts a loose constraint on the model parameters and all the models seem plausible. Future improvement of the data and discovery of more such UDGs will reveal the true nature of the galactic kinematics and boost model discrimination capability. In the present scenario, we focus on the analysis whether such UDGs can accommodate alternative gravity models without the need for DM or not. Constraints can also be put on the gravity models using alternative methods such as rotation curve or cosmological study. Additionally, tidal stability of UDGs in their host environment can provide strong test of alternative gravity theories \cite{asencio2022distribution}. Note, that these studies are also sensitive to the distance of the UDGs as the distance is crucial in estimating the baryonic mass density. For our study we take the mass density corresponding to the typical distance of DF2 and DF4 at $20$ Mpc. For the other claim of $13.2$ Mpc \cite{trujillo2019distance} the results would get modified. One may study the alternative gravity models including the uncertainty of the mass density models. These issues are beyond the scope of our present analysis. \\
Future discovery of similar galaxies will definitely build a stronger avenue for these alternative gravity models. Recently, a large number of UDGs were found in the Coma cluster \cite{vanDokkum:2016uwg,vanDokkum:2014cea,freundlich2022probing}. These UDGs in contrast to DF2 and DF4 are rich in DM as observed from their high rotational velocity. 
Studying these UDGs in light of alternative gravity models will also be interesting. 

\acknowledgements
 Sovan acknowledges the support of the Max Planck India Mobility Grant from the Max Planck Society, supporting the visit and stay at MPP during the project. Sovan has also received funding from DST-SERB projects CRG/2021/002961 and MTR/2021/000540. Sayan Chakrabarti would like to acknowledge the support from the DST-SERB research grant MTR/2022/000318. All the authors would like to acknowledge discussions and valuable inputs from Koushik Dutta, Tousif Islam, Sayak Dutta, Ashwani Rajan and Aritra Biswas. 

\nocite{*}
\bibliography{apssamp}

\end{document}